\documentclass[preprint, superscriptaddress]{revtex4-1}

\usepackage{graphicx, tabularx, subfigure, comment, multirow, rotating}
\usepackage[hidelinks]{hyperref}

\begin{document}
\preprint{R133}

\title{Enhanced charge ordering transition in doped CaFeO$_3$ through steric templating}

\author{Lai Jiang}
\affiliation{The Makineni Theoretical Laboratories, Department of Chemistry, University
of Pennsylvania, Philadelphia, PA 19104}
\author{Diomedes Saldana-Greco}
\affiliation{The Makineni Theoretical Laboratories, Department of Chemistry, University
of Pennsylvania, Philadelphia, PA 19104}
\author{Joseph T. Schick}
\affiliation{Department of Physics, Villanova University, Villanova, PA 19085}
\author{Andrew M. Rappe}
\email[]{rappe@sas.upenn.edu}
\affiliation{The Makineni Theoretical Laboratories, Department of Chemistry, University
of Pennsylvania, Philadelphia, PA 19104}
\date{\today}

\begin{abstract}

We  report density functional theory (DFT) investigation of $B$-site doped CaFeO$_3$, a prototypical
charge-ordered perovskite. At 290~K, CaFeO$_3$ undergoes a metal-insulator transition and a charge 
disproportionation reaction 2Fe$^{4+}$$\rightarrow$Fe$^{5+}$+Fe$^{3+}$. We observe that when Zr dopants occupy
a (001) layer, the band gap of the resulting solid solution increases to 0.93~eV due to a 2D Jahn-Teller type distortion,
where FeO$_6$ cages on the $xy$ plane elongate along $x$ and $y$ alternatively between neighboring Fe sites.
Furthermore, we show that the rock-salt ordering of the Fe$^{5+}$ and Fe$^{3+}$ cations can be enhanced when the
$B$-site dopants are arranged in a (111) plane due to a collective steric effect that facilitates the size discrepancy 
between the Fe$^{5+}$O$_6$ and Fe$^{3+}$O$_6$ octahedra and therefore gives rise to a larger band gap. The
enhanced charge disproportionation in these solid solutions is verified by rigorously calculating the oxidation states of 
the Fe cations with different octahedral cage sizes. We therefore predict that the corresponding transition temperature will
increase due to the enhanced charge ordering and larger band gap. 
The compositional, structural and electrical relationships exploited
in this paper can be extended to a variety of perovskites and non-perovskite oxides  providing guidance in structurally
 manipulating electrical properties of functional materials.

\end{abstract}

\maketitle

\section{Introduction}

The perovskite ($AB$O$_3$) family of materials has been paid considerable attention both in experimental and 
theoretical studies due to their flexible and coupled compositional, structural, electrical and magnetic properties
~\cite{Bellaiche00p5427, Saito04p84, Bilc06p147602}.
 Such flexibility arises from the structural building
blocks --- the corner-connected $B$O$_6$ octahedra, where $B$ is usually a transition metal.
 Typical structural variations from the cubic structure include rotation and tilting of
the octahedra~\cite{Glazer72p3384},
 off-centering of the $A$ and/or $B$ cations (pseudo Jahn-Teller effect)~\cite{Qi10p134113}, and expansion/contraction of the $B$O$_6$ octahedral cages
 ~\cite{Thonhauser06p2121061}.
While the first two distortions are ubiquitous, the cooperative octahedral breathing distortion is rather rare in perovskites with a 
single $B$ cation composition. Such breathing distortion, resulting from the alternation of elongation and contraction of the $B$-O bonds
between neighboring $B$O$_6$ cages, is usually concomitant with the charge ordering of the $B$ cations and a corresponding
metal-insulator transition. CaFeO$_3$ is a typical perovskite material exhibiting such charge ordering transition~\cite{Woodward00p844}.
 At room temperature,
the strong covalency in the Fe $e_g$ - O 2$p$ interaction leads to a $\sigma^*$ band and electron delocalization which
 gives rise to metallic conductivity in CaFeO$_3$. Near 290~K, a second-order
 metal-insulator transition (MIT) occurs which reduces the
 conductivity dramatically~\cite{Kawasaki98p1529}. The M\"{o}ssbauer spectrum of low temperature CaFeO$_3$ has revealed the
 presence of two chemically distinct Fe sites (with different hyperfine fields) present in equal proportion~\cite{Takano77p923}.
 This indicates that the Fe cations undergo charge disproportionation 
 2Fe$^{4+}$$\rightarrow$Fe$^{5+}$+Fe$^{3+}$ below the transition temperature. The origin of the charge
 ordering  transition is usually attributed to Mott insulator physics, where
 the carriers are localized by strong electron-lattice interactions~\cite{Millis98p147, Takano77p923, Ghosh05p245110,
 Woodward00p844}. More recently, it has been debated whether the difference in charge state resides 
 on the $B$ cations or as holes  in the oxygen $2p$ orbitals~\cite{Yang05p10A312, Akao03p156405,
 Mizokawa00p11263}, and several computational studies showed that the magnetic configuration, in addition
 to structural changes, plays a vital role in stabilizing the charge ordered state in CaFeO$_3$~\cite{
 Mizokawa98p1320, Ma11p224115, Cammarata12p195144}.
 Nevertheless, the amplitude of the
 cooperative breathing mode is a key indicator of the magnitude of  electron trapping and band gap
 opening in MIT. Conversely because the MIT is sensitive to lattice distortion,  structural manipulation such as cation 
 doping and epitaxial strain
 can be exploited to control the electrical properties
 of this family of oxide systems.
 
 In this study, we examine the structural and electrical properties in $B$-cation doped CaFeO$_3$ with 
 density functional theory (DFT). Various dopant cations, concentrations, and arrangements
  have been tested. Dopants of different sizes are tested, and alignments of pairs of dopants along
  different crystallographic planes are examined. To confirm the presence  of 
charge ordering in (111) doped CaFeO$_3$, we also carried out rigorous oxidation state calculations for Fe 
cations in different octahedral cages based on their wave function topologies~\cite{Jiang12p166403}. 
Through examination of these model systems, we assess the extent to which the structure-coupled 
electronic transition in doped oxide materials like CaFeO$_3$  can be influenced via doping to enhance
band gap tunability, which in turn controls the MIT temperature.

\section{Methodology}

Our DFT calculations are performed using the norm-conserving nonlocal pseudopotential plane-wave method
~\cite{Payne92p1045}. The pseudopotentials~\cite{Rappe90p1227} are generated by the \textsc{Opium} 
package~\cite{OPIUM} with a 50~Ry plane-wave energy cutoff.  
Calculations are performed with the \textsc{Quantum-Espresso} package~\cite{Giannozzi09p395502} using the
local density approximation~\cite{Perdew81p5048} with the rotationally invariant effective Hubbard $U$ 
correction~\cite{Johnson98p15548} of 4~eV on
the Fe $d$ orbitals~\cite{Fang01p180407, Cammarata12p195144} for the exchange-correlation functional.
In case of Ni and Ce doping, we applied $U$ = 4.6~eV~\cite{Cococcioni05p035105}
 and $U$ = 5~eV~\cite{Loschen07p035115} for Ni $d$ and Ce $f$, respectively.
 Calculations are
performed on a $4\times4\times4$ Monkhorst-Pack $k$-point grid~\cite{Monkhorst76p5188} with electronic
energy convergence of $1\times10^{-8}$~Ry, force
convergence threshold of $2\times10^{-4}$~Ry/\AA, and pressure convergence threshold of 0.5 kbar. For
polarization calculations a $4\times6\times12$ $k$-point grid is used, where the densely sampled direction
is permuted in order to obtain all three polarization components. Different spin orderings for pure
CaFeO$_3$ are tested to find the magnetic ground state, and subsequent solid solution calculations all start
with that magnetic ground state.

\section{Results and discussion}

\subsection{Ground state of CaFeO$_3$ and CaZrO$_3$}
\begin{figure}[htp]
\centering
\includegraphics[width=\textwidth]{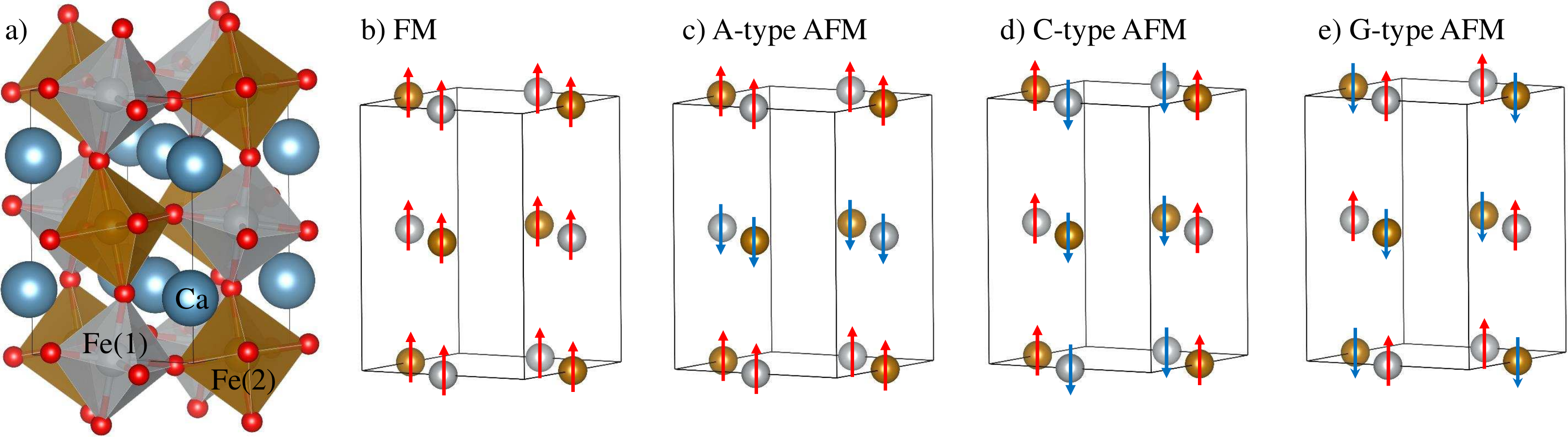}
\caption{(a) Low temperature $P2_1/n$ structure of CaFeO$_3$ with two symmetry-distinct Fe cation sites color coded. The spin ordering of the Fe cations are (b) ferromagnetic (FM), (c) A-type anti-ferromagnetic (AFM), 
(d) C-type anti-ferromagnetic and (e) G-type anti-ferromagnetic.}
\label{fig:CFO}
\end{figure}

To identify the correct spin ordering in pure CaFeO$_3$, we performed relaxations on both high temperature 
metallic orthorhombic $Pbnm$~\cite{Takano77p923, Ghosh05p245110, Woodward00p844, Kanamaru70p257} and 
low temperature semiconducting monoclinic $P2_1/n$~\cite{Saha-Dasgupta05p045143} structures with common
magnetic orderings commensurate with the $2\times2\times2$ supercells, as shown in Fig.~\ref{fig:CFO}
(note that diamagnetic (DM) ordering is not included in the figure).

\begin{table}[htp]
\caption{Calculated total energy $E$, atomic magnetization $m_1$ and $m_2$ for the two Fe sites, 
total magnetization per five-atom formula unit  $M$ and FeO$_6$ octahedral volume $V_1$ and $V_2$ for
the two Fe sites of relaxed CaFeO$_3$ with different starting structures and magnetic orderings.}
\begin{tabularx}{\textwidth}{ X  X X X X X X X}
  \hline\hline
  & & $E$ (eV) & $m_1$ ($\mu_\mathrm{B}$) & $m_2$ ($\mu_\mathrm{B}$) & $M$ ($\mu_\mathrm{B}$)
  & $V_1$ (\AA$^3$) & $V_2$ (\AA$^3$) \\
  \hline
  \multirow{5}{*}{\begin{sideways}$Pbnm$\end{sideways}} & DM & 6.64 & N/A & N/A & N/A & 8.40 & 8.40 \\
  & FM & 0.07 & 3.38 & 3.38 & 4.00 & 8.94 & 8.94 \\ 
  & A-AFM & 0.43 & 3.29 & -3.29 & 0.00 & 9.00 & 9.00 \\
  & C-AFM & 0.58 & 3.24 & -3.24 & 0.00 & 8.92 & 8.92 \\
  & G-AFM & 0.96 & 3.44 & -3.51 & 0.06 &  9.52 & 8.47\\
  \hline
  \multirow{5}{*}{\begin{sideways}$P2_1/n$\end{sideways}} & DM & 6.64 & N/A & N/A & N/A & 8.40 & 8.40  \\
  & FM & 0 & 3.13 & 3.60 & 4.00 & 9.16 & 8.75 \\
  & A-AFM & 0.32 & 3.69 & -3.69 & 0.00 & 9.57 & 8.40 \\
  & C-AFM & 0.58 & 3.24 & -3.24 & 0.00 & 8.92 & 8.92 \\
  & G-AFM & 0.81 & 3.85 & -2.39 & 1.00 & 10.05 & 8.13 \\
  \hline\hline
\end{tabularx}
\label{tab:CFO}
\end{table}

\begin{figure}[htp]
\centering
\includegraphics[width=\textwidth]{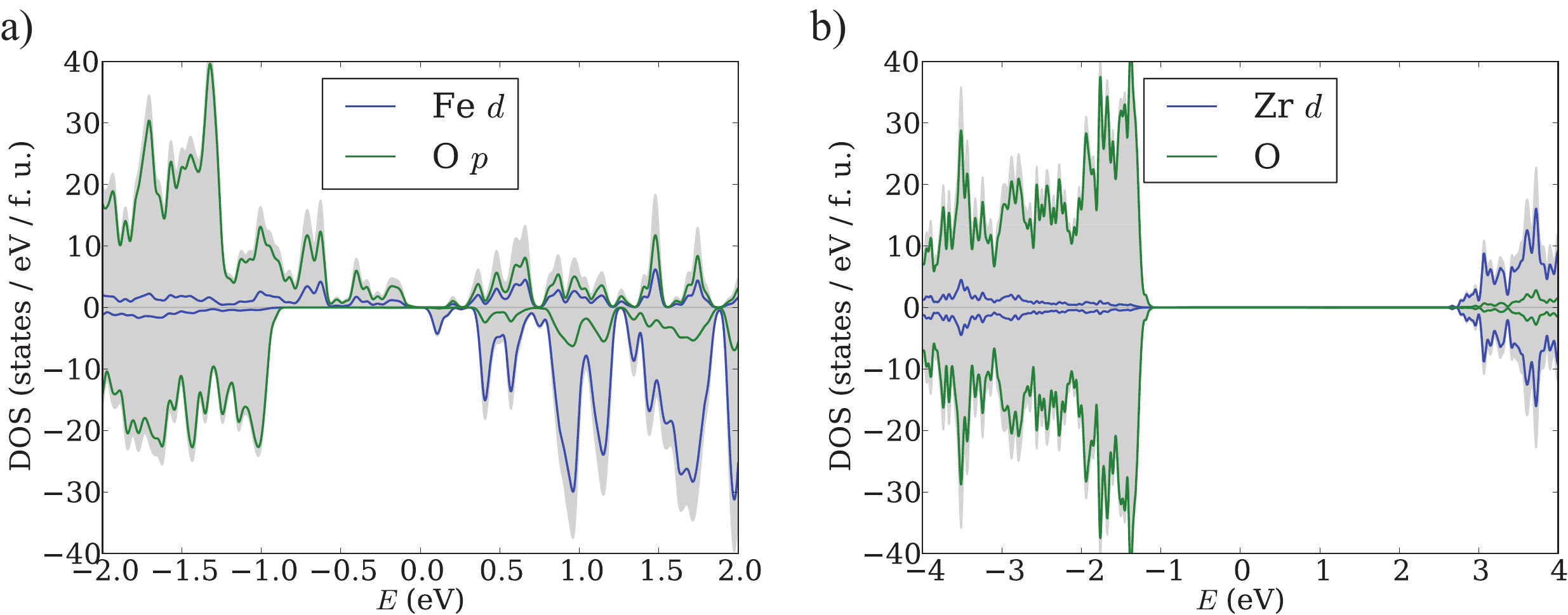}
\caption{Projected density of states of (a) CaFeO$_3$ (left) and (b) CaZrO$_3$.}
\label{fig:PDOS}
\end{figure}

From the results in Table.~\ref{tab:CFO}, we can see that both high temperature and low temperature ferromagnetic CaFeO$_3$
relax to the ferromagnetic ground states. Note that  an additional magnetic phase transition is experimentally observed for CaFeO$_3$ at 15~K, 
where it adopts an incommensurate magnetic 
structure with a modulation vector [$\delta$, 0, $\delta$] ($\delta\approx0.32$, 
and reciprocal lattice vectors as basis)~\cite{Woodward00p844}. Since DFT calculates 0~K
 internal energy, the ferromagnetic ground state represents a reasonable
approximation of the spin-spin interactions within a unit cell given the relatively long spin wave length and low
experimental crossover temperature to FM.
The volumes of the two
FeO$_6$ cages are equivalent in the high temperature metallic phase, as expected. The low temperature ground
state has a cage size difference $\Delta V =0.41 $\AA$^3$, indicating some degree of charge ordering. However, the projected density of states (PDOS) of the
$P2_1/n$ ground state CaFeO$_3$ in Fig.~\ref{fig:PDOS}a shows that although there are separate gaps in each 
spin channel, the valence band edge in the majority spin touches the conduction band edge in 
the minority spin, resulting in
 zero total gap. The absence of band gap and the weak charge ordering is a result of the underestimation of band gaps in DFT~\cite{
Kohn65pA1133} due to its unphysical electron delocalization, 
 This result is common in Mott insulators with partially filled $d$ orbitals  and is in agreement with another 
DFT study of CaFeO$_3$~\cite{Yang05p10A312}. Even though DFT does not predict the correct electronic ground state of CaFeO$_3$, it is 
however indicative of the sensitive nature of the CaFeO$_3$ band gap as it can be easily influenced when Fe $d$ orbital filling 
is varied by structural or other perturbations. Moreover, the different $\Delta V$ between high temperature and 
low temperature of FeO$_6$ ground state suggests that the structural aspect of the MIT can be 
modeled reasonably well by DFT. 

For comparison, we  calculated the PDOS of CaZrO$_3$ relaxed from experimental $Pbmn$ structure~\cite{Levin03p170}, shown also
in Fig.~\ref{fig:PDOS}b. Because Zr$^{4+}$ has empty $4d$ orbitals, the fraction of Zr $d$ states in the valence
 band is negligible compared to O $p$ states, and a wide charge-transfer gap of 3.82~eV occurs. Since we 
 expect to exploit the size effect of the dopants like Zr to influence the electronic property of CaFeO$_3$,
we expect that the nature of Fe spin-spin interaction is not greatly affected by doping. Therefore
 in the following study we continue to use FM as the starting magnetic configuration for relaxations of the
 doped materials.

\subsection{CaFeO$_3$-CaZrO$_3$ solid solutions with $2\times2\times2$ super cell}

\begin{figure}[htp]
\centering
\includegraphics[width=\textwidth]{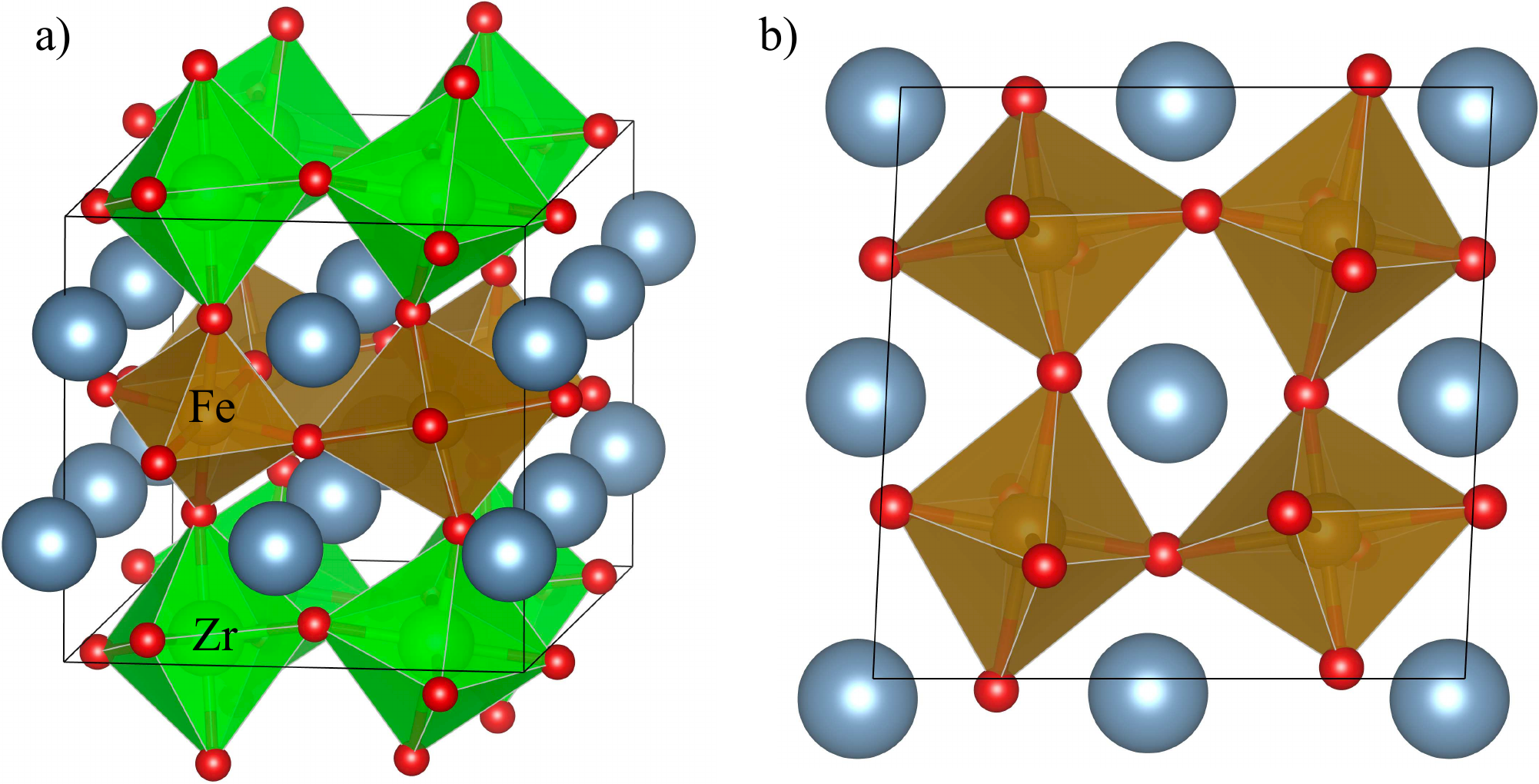}
\caption{(a) Crystal structure of Ca(Fe$_{1/2}$Zr$_{1/2}$)O$_3$ and (b) top view of the ZrO$_2$ layer showing
the 2D Jahn-Teller type distortion.}
\label{fig:JT}
\end{figure}

To test how Zr doping influences the structural and electrical properties of CaFeO$_3$, we performed relaxations
and subsequent band gap calculations of CaFeO$_3$-CaZrO$_3$ solid solutions. We employ a $2\times2\times2$ 
super cell and explore all possible $B$-site cation combinations. All the solid
solutions tested turn out to be metallic except one, which has a  gap of 0.93~eV. The insulating solid
solution has a cation arrangement with four Zr cations  on the (001) plane, 
making it a layered 
structure along [001]. Interestingly, instead of a breathing-mode charge disproportionation,
 this structure has a 2D Jahn-Teller type distortion, and all four Fe cations are in the same
chemical environment. As shown in Fig.~\ref{fig:JT}, the Fe-O bond lengths in the $xy$ plane are 2.16~\AA\ 
and 1.83~\AA\  in each FeO$_6$, with the orientation alternating between neighbors. The shorter Fe-O bond length is 
essentially the same as that in high temperature CaFeO$_3$. The consequence of the addition of larger Zr$^{4+}$ cations ($r=0.72$~\AA) compared to
 Fe$^{4+}$ ($r=0.59$~\AA) is that when Zr cations 
occupy an entire (001) plane the in-plane lattice is expanded from 3.70~\AA\ to 3.88~\AA,  
elongating the Fe-O bonds  and enabling the 2D Jahn-Teller type distortion. 

\begin{figure}[htp]
\centering
\includegraphics[width=\textwidth]{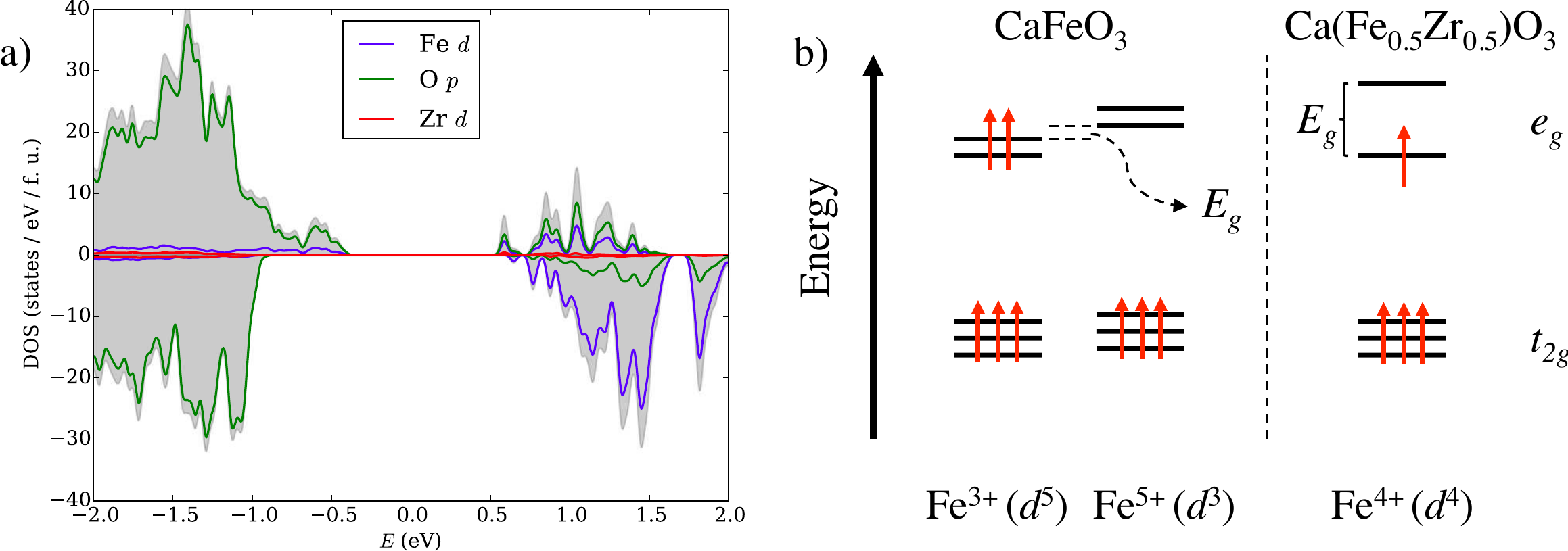}
\caption{(a) PDOS of of Ca(Fe$_{1/2}$Zr$_{1/2}$)O$_3$ and (b) illustration of band gap formation in
CaFeO$_3$ (left) and Ca(Fe$_{1/2}$Zr$_{1/2}$)O$_3$ (right).
}
\label{fig:CFZO_PDOS}
\end{figure}

From the PDOS of the solid solution in Fig.~\ref{fig:CFZO_PDOS}a we can see that like pure CaFeO$_3$ both
the valence and the conduction edges are of Fe $3d$ and O $2p$ characters, with virtually no Zr contribution.
In charge ordering MIT, the delocalized electrons on Fe$^{4+}$ transfer to neighboring
Fe$^{4+}$, making Fe$^{3+}$/Fe$^{5+}$ pairs with the valence and conduction bands located on 
different cations, concomitant with FeO$_6$ cage size changes. The band gap in case of charge ordering therefore
depends on the energy difference between the $e_g$ orbitals in Fe$^{3+}$ and Fe$^{5+}$, which in turn
is affected by the crystal field splitting energy caused by the oxygen ligands. On the other hand, 
as illustrated in Fig.~\ref{fig:CFZO_PDOS}b,
the solid solution band gap is caused by the removal of degeneracy in the $e_g$ orbitals and is 
controlled by the difference in energy between the two $e_g$ orbitals on the same Fe$^{4+}$ cation.
Since the $e_g$ gap splitting is a result of the Fe-O bond length difference, it is easier to tune by applying
either chemical pressure or biaxial strain to change the in-plane lattice constant, whereas the charge
ordering mechanism requires the control of individual FeO$_6$ octahedral sizes to change the relative
energy of $e_g$ orbitals between two Fe sites.  Nevertheless, this CaFeO$_3$-CaZrO$_3$ solid
solution demonstrates that when arranged in a particular way, in this case on (001) plane, the size effect
of the large Zr cation can cause a cooperative steric effect on the structure and affect the electrical properties of 
CaFeO$_3$, opening up the band gap via a completely different mechanism.

\subsection{CaFeO$_3$ with dopants on the (111) plane}

\begin{figure}[htp]
\centering
\includegraphics[width=\textwidth]{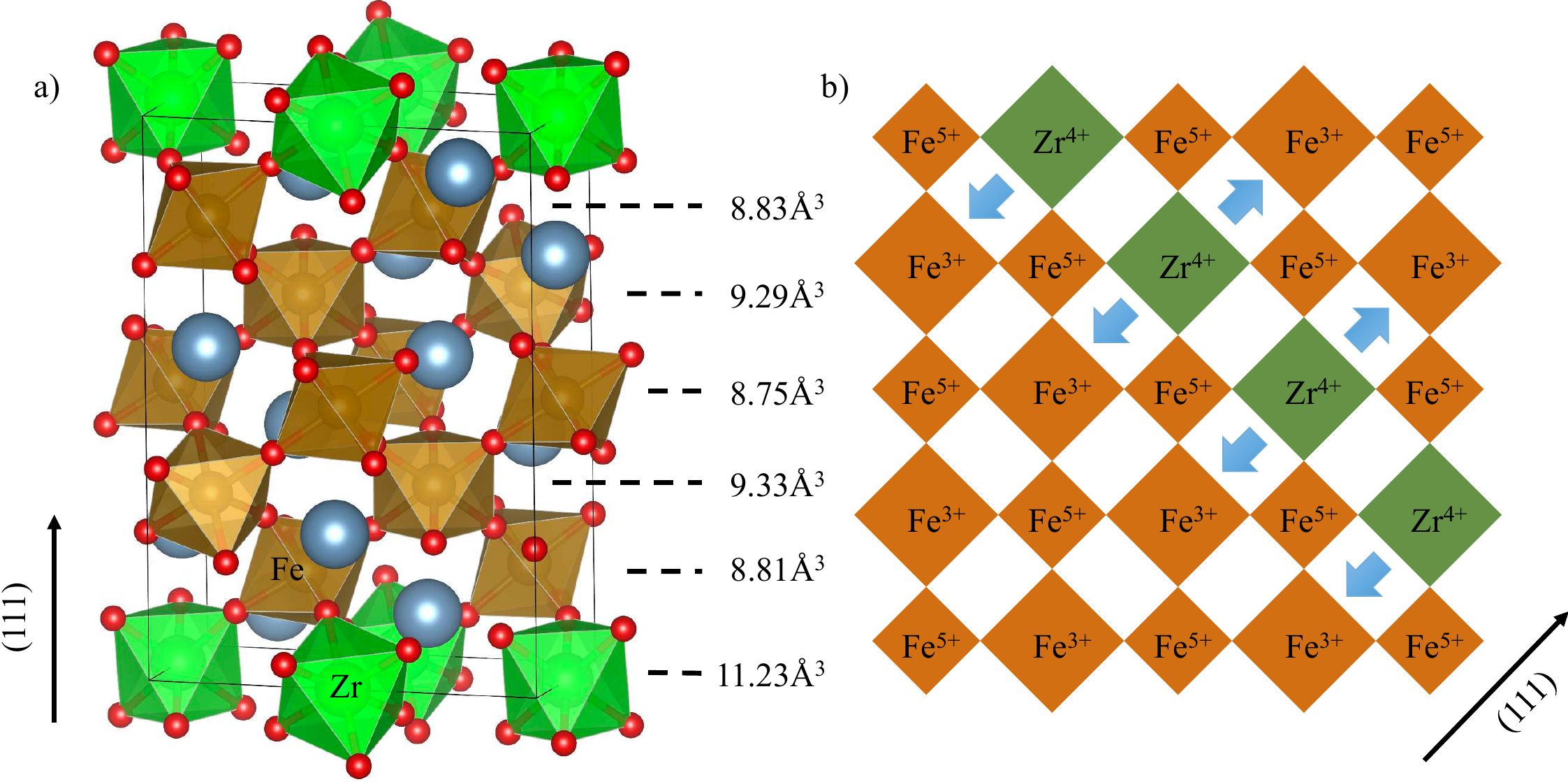}
\caption{(a) Crystal structure of $\sqrt{2}\times\sqrt{6}\times2\sqrt{3}$ CaFeO$_3$ supercell doped with one layer of
Zr on the (111) plane. The average $B$O$_6$ octahedron size is listed on the side. (b) Schematic of how a layer of dopants with
larger ionic radius exerts a cooperative size effect on the neighboring layers and enhances the existing charge ordering.}
\label{fig:CFZO_111}
\end{figure}

As discussed in the previous section, simply doping Zr into $2\times2\times2$ CaFeO$_3$ does not increase the
band gap except for one case where 2D Jahn-Teller instead of breathing mode serves to make the system insulating. 
In a perovskite system with rock-salt ordered alternating $B$ cations, such as CaFeO$_3$, one $B$ cation type
occupies entire (111) planes and the other type occupies its neighbors in all directions. Since Zr cation is larger than Fe cation,
to fully utilize its steric effect to distinguish  Fe$^{3+}$ from Fe$^{5+}$, it follows that Zr should replace a full Fe$^{3+}$ 
plane to increase the $B$O$_6$ size on the plane, maximizing its utility by enhancing  the cage size difference. A schematic of the (111) doping
strategy and the influence of the dopants on their neighboring planes is shown in Fig.~\ref{fig:CFZO_111}b. 

Following this logic, we perform calculations with $\sqrt{2}\times\sqrt{6}\times2\sqrt{3}$ CaFeO$_3$ super cell, which has
 six (111) FeO$_2$ layers stacked perpendicularly, as the parent material. One layer of Fe's is replaced with  Zr's and the structural is
 relaxed. The final structure is 
shown in Fig.~\ref{fig:CFZO_111}a, as well as the average $B$O$_6$ cage size of each layer. Clearly the introduction of a (111) Zr layer
drives the charge disproportionation of Fe$^{4+}$ by exerting chemical pressure on both sides of the layer and favoring the FeO$_6$ 
on the two adjacent planes to be smaller and become Fe$^{5+}$. The second next neighboring layers in turn have more room to expand and favor larger Fe$^{3+}$.
The size difference between the largest and the smallest FeO$_6$ cages ($\Delta V =0.58$\AA$^3$) in this structure is 
an enhancement compared to pure CaFeO$_3$ $\Delta (V=0.41$\AA$^3$), which
suggests the presence of a stronger charge ordering and a wider band gap. However electronic structure calculation shows that this 
solid material is metallic as well. The reason that the seemingly more charge ordered system still does not possess a gap can be attributed to the supercell
 employed. By using a unit cell with six (111) layers and replacing only one layer of Fe with Zr, structurally the remaining five Fe layers
are disturbed by the large Zr layer as expected. However the charge disproportionation reaction 
2Fe$^{4+}$$\rightarrow$Fe$^{5+}$+Fe$^{3+}$ cannot proceed to completion, because it requires an even number of Fe layers.
Therefore with one layer of dopants there will always be Fe$^{4+}$ ``leftovers'' that render the whole system metallic.

To resolve the issue of odd number of Fe layers we introduce another layer of +4 dopants with smaller ionic radius than Fe.
For simplicity we denote a solid solution in this case by listing the $B$ cations in each of its six (111) layers, with dopant elements in
bold. For example,
the previously discussed one layer Zr-doped solid solution would be denoted as \textbf{Zr}FeFeFeFeFe.
 The
presence of two dopant layers provides both positive and negative chemical pressure to expand Fe$^{3+}$ and contract Fe$^{5+}$.
These two dopant layers are separated by an even number of Fe layers so that the FeO$_6$ size alternation is enhanced. An odd number
of Fe layers in between the dopant layers would disrupt and impede the size modulation period.
Relaxations are performed on \textbf{ZrNi}FeFeFeFe and \textbf{Zr}FeFe\textbf{Ni}FeFe, as well as \textbf{CeNi}FeFeFeFe and
 \textbf{Ce}FeFe\textbf{Ni}FeFe. The average
FeO$_6$ cage size per layer is listed in Table~\ref{tab:size_111}, along with the maximum cage size difference $\Delta V$
 and the corresponding band gap of each solid solution. It can be seen that with only Zr as dopant, the $\Delta V$ is significantly smaller
 than those with two layers of dopants, and $\Delta V$ correlates with the band gap. The Ce-containing solid solutions have a larger $\Delta V$
 compared to the Zr-containing ones, due to the larger size of Ce. It also shows that when the larger dopant and the smaller dopant layers are
 adjacent, the resulting $\Delta V$ is larger than when they are two Fe layers apart, this is due to the lack of symmetry of the former 
 configuration where the absence of a mirror plane perpendicular to the $z$ axis allows for the FeO$_6$ close to the dopant layers to further
 expand or contract compared to the ones that are not neighbors of the dopant layers. In the latter configuration, symmetry guarantees that
 octahedra on either side of the dopant layer are deformed equally.

\begin{table}[htp]
\caption{Properties of CaFeO$_3$ doped on the (111) plane. $V_1$ through $V_6$ are average FeO$_6$ volumes in \AA$^3$, 
where the largest and 
the smallest cages in each solid solution are in bold. $\Delta V$ is the size difference between the largest and smallest volumes and $E_g$ is the band gap 
of the corresponding material in eV.}
\begin{tabularx}{\textwidth}{X X X X X X X X X X}
  \hline\hline
  \multicolumn{2}{l}{(111) Layers} &  $V_1$ &  $V_2$ & $V_3$ & $V_4$ & $V_5$ & $V_6$ & $\Delta V$ & $E_g$ \\
  \hline
  \textbf{Zr}FeFeFeFeFe & & Zr & 8.81 & \textbf{9.33} & \textbf{8.75} & 9.29 & 8.83 & 0.58 & 0 \\
  \textbf{ZrNi}FeFeFeFe & & Zr & Ni & 9.52 & 8.53 & \textbf{9.59} & \textbf{8.27} & 1.32 & 0.49 \\
  \textbf{Zr}FeFe\textbf{Ni}FeFe & & Zr & 8.32 & 9.43 & Ni & \textbf{9.45} & \textbf{8.28} & 1.17 & 0.11 \\
  \textbf{CeNi}FeFeFeFe & & Ce & Ni & \textbf{9.77} & 8.61 & 9.69 & \textbf{8.29} & 1.48 & 0.83 \\
  \textbf{Ce}FeFe\textbf{Ni}FeFe & & Ce & \textbf{8.39} & \textbf{9.75} & Ni & 9.73 & 8.41 & 1.36 & 0.53 \\
  \hline\hline
\end{tabularx}
\label{tab:size_111}
\end{table}

\begin{figure}[htp]
\centering
\includegraphics[width=\textwidth]{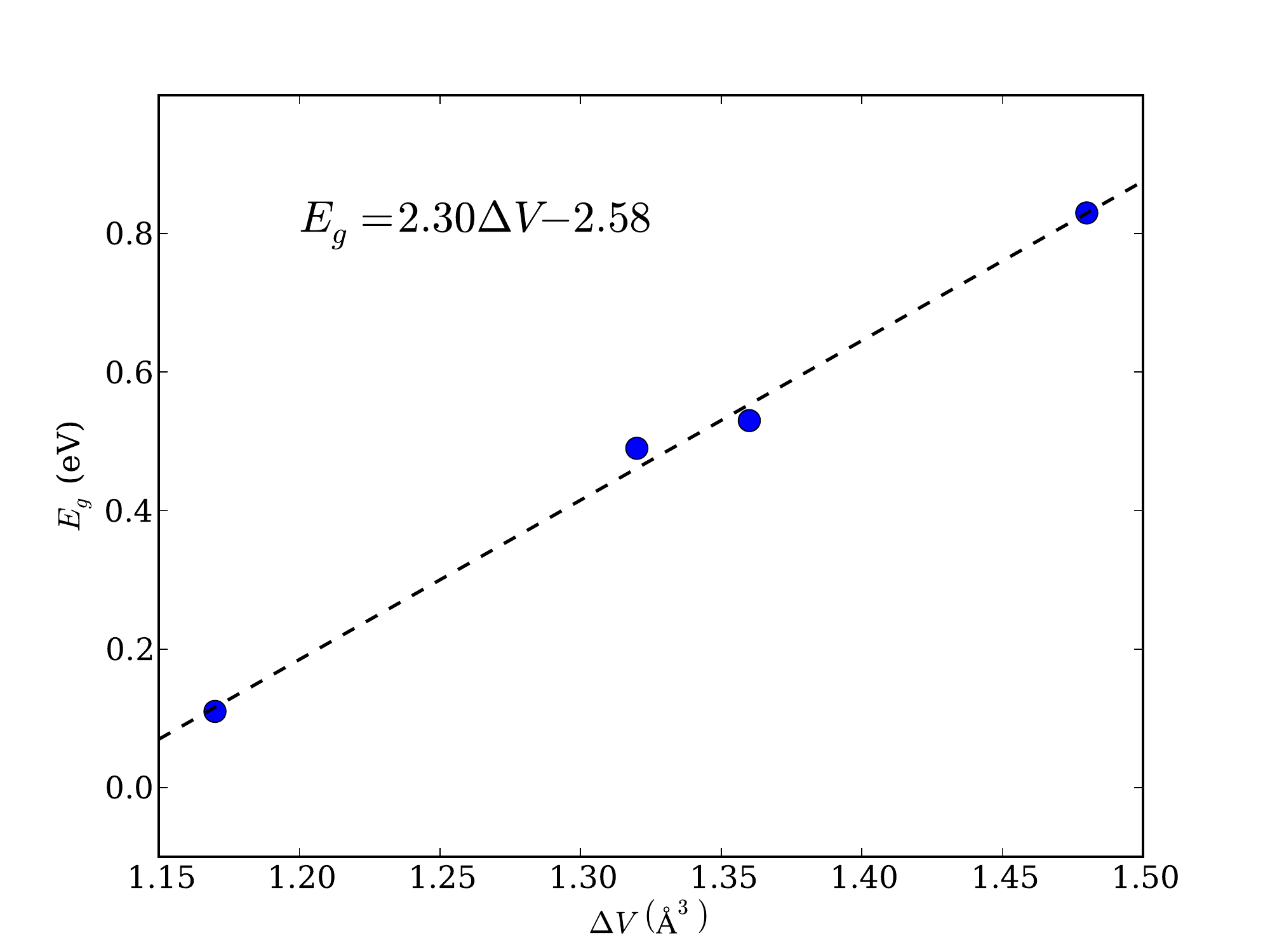}
\caption{Band gap $E_g$ of the (111) doped CaFeO$_3$ solid solutions increases with the corresponding maximum FeO$_6$ size
 difference $\Delta V$. This gives an effective chemical pressure on the band gap of 2.30~eV/\AA$^3$ or 370~GPa.}
\label{fig:gap}
\end{figure}

From Fig.~\ref{fig:gap} we can see that with increasing difference in FeO$_6$ size, the band gap of the corresponding solid solutions
increases accordingly. This relationship demonstrates the coupling between structural  and electrical properties as  larger
FeO$_6$ size discrepancy indicates stronger and more complete charge disproportionation. As illustrated in Fig.~\ref{fig:CFZO_PDOS}b, when charge
ordering is the band gap opening mechanism, the gap size depends on the crystal field splitting energy difference between Fe$^{3+}$ and Fe$^{5+}$.
A larger FeO$_6$ cage size difference means that the O $2p$ - Fe $3d$ repulsion difference is also larger between
the two Fe sites. This causes the energy difference of the $e_g$ orbitals in the two sites to increase and the band gap to increase as well.
 Using linear regression we estimate that the chemical pressure
 excerted on the band gap by the volume difference in this type of solid solutions is quite large at about 370~GPa, in accordance with the effective 
 band gap tuning. Since the transition to metal occurs when thermally activated electrons have enough energy to cross the band gap
 and flow between the two Fe sites to make them indistinguishable, we believe that by (111) doping the MIT temperature of CaFeO$_3$
 can be increased, making devices based on it more operable at room temperature.
 
\begin{figure}[htp]
\centering
\includegraphics[width=\textwidth]{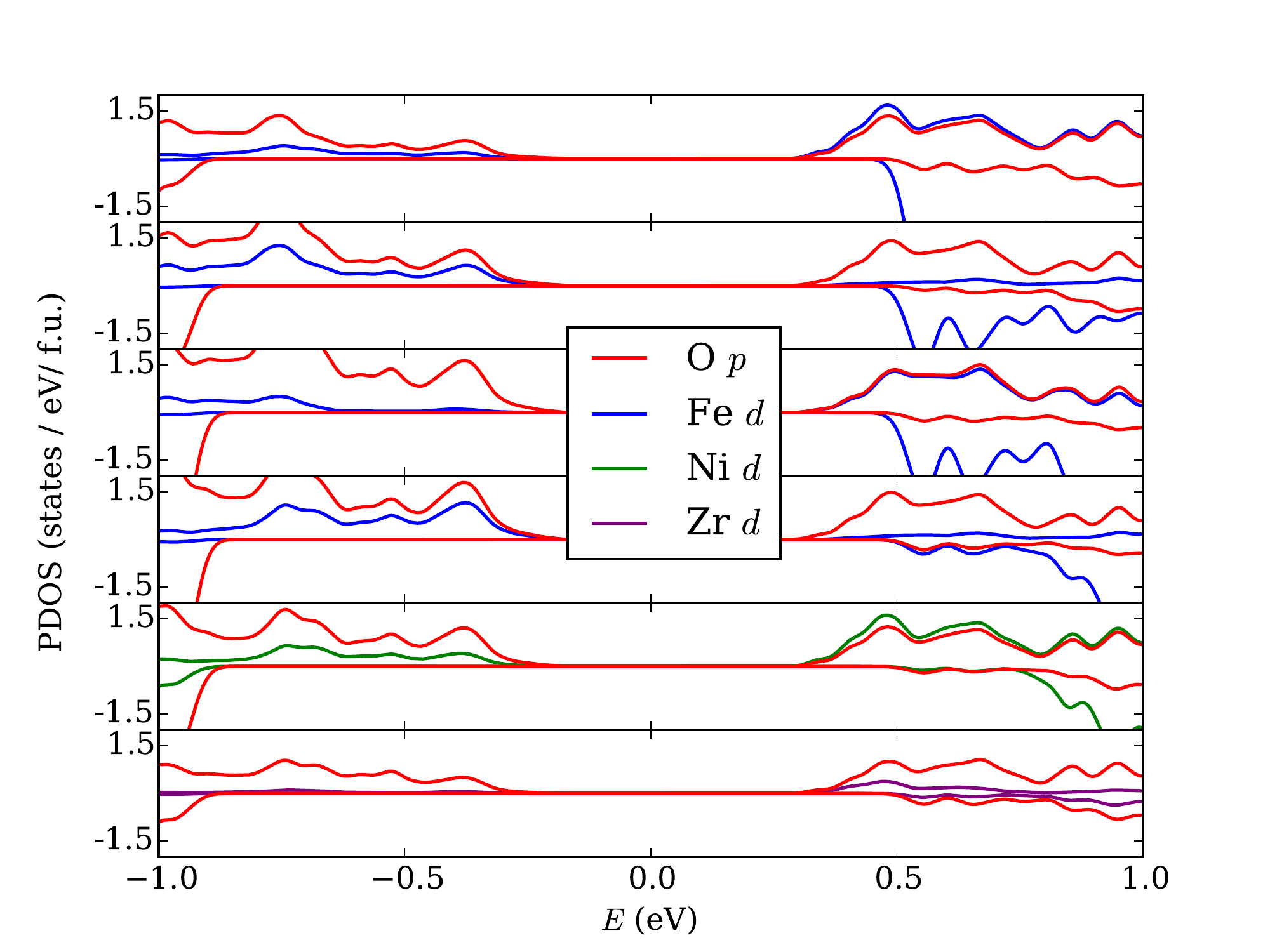}
\caption{Layer resolved projected density of states of \textbf{ZrNi}FeFeFeFe. Each of the six panels represents a layer of 
Ca$B$O$_3$, and the relative position of the panels corresponds to the that of the six layers in the crystal.}
\label{fig:layered_PDOS}
\end{figure}

To investigate the  layered nature of the solid solutions, we use \textbf{ZrNi}FeFeFeFe as an example and plot  
the projected density of states of it in Fig.~\ref{fig:layered_PDOS} in a layer resolved fashion. Each of the six panels in 
Fig.~\ref{fig:layered_PDOS} represents a layer of 
Ca$B$O$_3$, and the relative position of the panels corresponds to the that of the six layers in the crystal. It can be seen clearly
that for the four layers containing Fe ions, the first and the third layers have more majority spin Fe $d$  in the valence
band, while the second and fourth layers have more majority spin Fe $d$ in the valence band. This difference is consistent with the
fact that Fe$^{3+}$ has more filled $d$ orbitals than Fe$^{5+}$ and supports our prediction that the doubly doped (111) layered CaFeO$_3$
has an enhanced charge ordering due to the strong modulation of the FeO$_6$ cage volume.

\begin{figure}[htp]
\centering
\includegraphics[width=\textwidth]{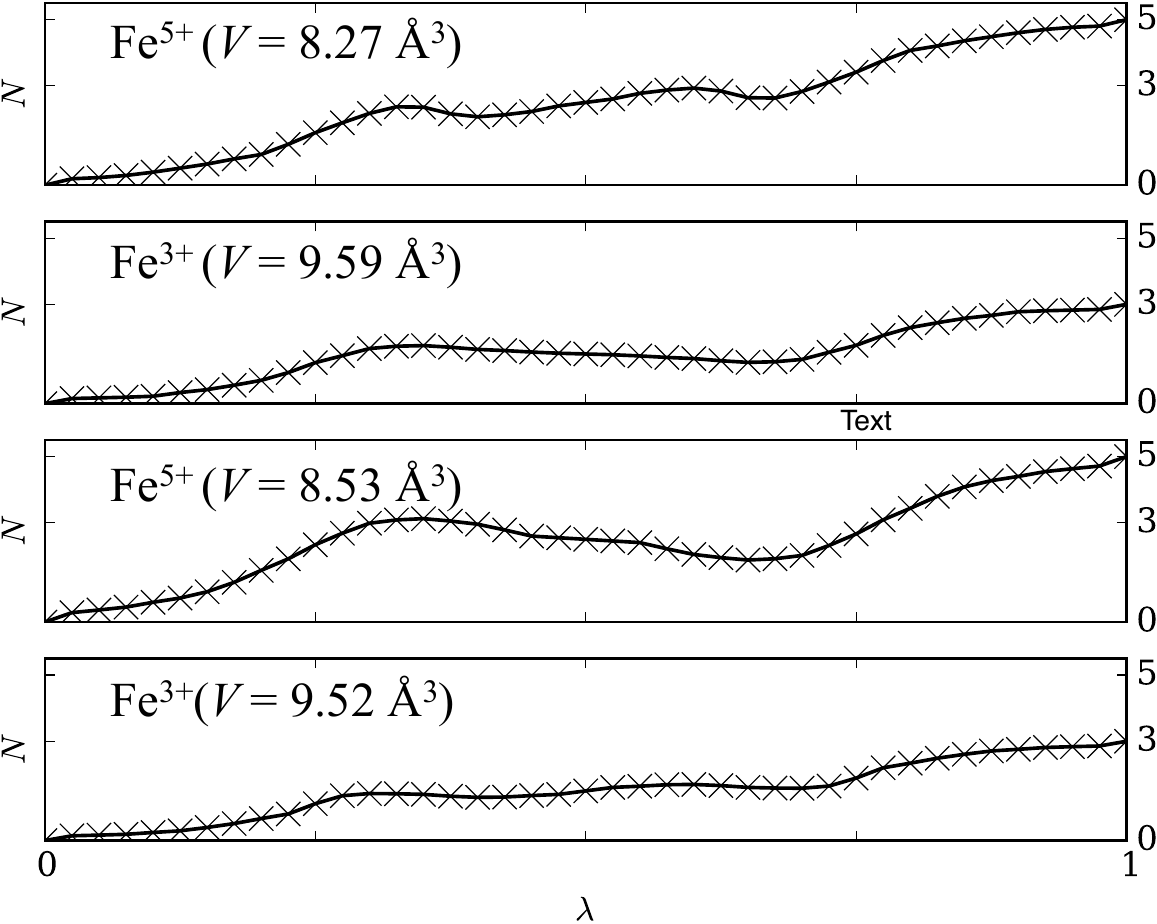}
\caption{Oxidation state $N$ of the four Fe cations. $\lambda$ denotes the reaction coordinate of moving the Fe ion
sublattice to the neighboring, cell
and the change in $N=\Delta\frac{\vec{P}\cdot\vec{R}}{\vec{R^2}}$ from $\lambda=0$ to $\lambda=1$ corresponds to the oxidation state of that Fe ion.}
\label{fig:ox}
\end{figure}

To further verify the charge disproportionation mechanism, 
we performed oxidation state calculations of the Fe cations in \textbf{ZrNi}FeFeFeFe. We employed
an unambiguous oxidation state definition~\cite{Jiang12p166403} based on wave function topology,  whereby moving a target ion to its image site in an adjacent  cell
through an insulating path and calculating the polarization change during the process, the number of electrons that accompany the 
moving nucleus can be calculated. The oxidation state obtained this way is guaranteed to be an integer  and is unique for an
atom in a given chemical environment, not dependent on other factors such as charge partitioning or the choice of orbital basis.
 In Fig.~\ref{fig:ox}
we show how the quantity $N = \Delta\vec{P}\cdot\vec{R}/\vec{R}^2$ changes as each Fe cation is moved along an insulating path to
the next cell, which is equivalent to the oxidation state of the cation. The two Fe ions with larger cages are 
confirmed to be Fe$^{3+}$ and the
ones with smaller cages are Fe$^{5+}$. This proves that charge ordering occurs in this material and causes band gap opening. 
Note that the oxidation states calculated are not directly related to the charges localized around the Fe sites, which has been
been shown to change insignificantly upon oxidation reaction in some cases~\cite{Sit11p12136}. In fact the Bader 
charge~\cite{Bader90} of the Fe cations are 1.76 and 1.73 for the smaller and larger FeO$_6$ cages, respectively,
which shows minuscule differences between Fe sites that  are in significantly different chemical environments in terms of oxygen 
ligand attraction. 

The (111)
doping strategy shows that the size difference between Fe$^{3+}$ and Fe$^{5+}$ can be exploited and reinforced by selectively replacing layers
of  Fe$^{3+}$ with Fe$^{5+}$ with atoms of even larger or smaller size, respectively, to enhance  charge ordering and the insulating
character of the CaFeO$_3$ system.

\section{Conclusions}

We have demonstrated that for prototypical charge ordering perovskite CaFeO$_3$, the band gap of the insulating state can
be engineered by $B$-site cation doping and structural manipulation. For the dopant atoms to exert significant influence on the parent material, it is 
favorable to arrange them in a way that  their size effects are cooperative and synergistic,  producing a collective steric effect
and greatly altering the structural and electrical properties. When doped on the (001) plane with larger Zr cations, the in-plane lattice constant
expands and
supports a 2D Jahn-Teller type distortion, where each FeO$_6$ has two distinct Fe-O bond length in the $xy$ plane. Such distortion removes
the degeneracy of the two $e_g$ orbitals on each Fe$^{4+}$ and opens up a band gap (not caused by charge ordering) of 0.93~eV. On
the other hand, to enhance the weak charge ordering in pure CaFeO$_3$, we discovered that including two types of 
dopants on the (111) plane can increase the FeO$_6$ cage size difference and enhance the charge ordering. Using
 Zr or Ce to
replace the larger Fe$^{3+}$ and Ni to replace the smaller Fe$^{5+}$ increases the band gap up to 0.83~eV. The degree of 
charge ordering  is closely related to the magnitude of FeO$_6$ cage size difference.
We used the rigorous definition
of oxidation state to verify that in the latter case the band gap opening mechanism is indeed charge disproportionation, as the oxidation states
of the larger and smaller Fe cations are calculated to be +3 and +5, respectively. Our results show that the structural and electrical properties of
CaFeO$_3$ are  coupled, and simple steric effects can enhance charge ordering transition and alter the band gap of the material greatly
 when the dopant atoms are placed to act cooperatively. Lastly by enhancing the charge ordering via doping, we predict that the MIT
 temperature of CaFeO$_3$ can also be increased to a temperature more suitable for practical device operation.

\begin{acknowledgments}

L. J. was supported by the Air Force Office of Scientific Research under Grant No. FA9550-10-1-0248.
D. S. G. was supported by the Department of Energy Office of Basic Energy Sciences under Grant No.~DE-FG02-07ER15920.
J. T. S. was supported by a sabbatical granted by Villanova University.
A. M. R. was supported by the Office of Naval Research under Grant No.~N00014-12-1-1033.
Computational support was provided by the High Performance Computing Modernization Office of the Department of Defense,
and the National Energy Research Scientific Computing Center of the Department of Energy.

\end{acknowledgments}

%

\end{document}